\documentclass[
  english,
  sigconf,
  letterpaper,
  ]{acmart}

\usepackage{paralist}

\usepackage{listings}

\usepackage{makecell}
\usepackage{multirow}

\usepackage{subcaption}

\usepackage[nopostdot, nomain, acronyms]{glossaries}
\glsdisablehyper{}
\setacronymstyle{long-short}
\newacronym{dom}{DOM}{Document Object Model}
\newacronym{url}{URL}{Uniform Resource Locator}
\newacronym{html}{HTML}{HyperText Markup Language}
\newacronym{css}{CSS}{Cascading Style Sheets}
\newacronym{js}{JS}{JavaScript}
\newacronym{json}{JSON}{JavaScript Object Notation}
\newacronym{idl}{IDL}{Interface Description Language}
\newacronym{gif}{GIF}{Graphics Interchange Format}
\newacronym{http}{HTTP}{HyperText Transfer Protocol}
\newacronym{api}{API}{Application Programming Interface}
\newacronym{ui}{UI}{User Interface}
\newacronym{seo}{SEO}{Search Engine Optimization}
\newacronym{waiaria}{WAI-ARIA}{Web Accessibility Initiative --- Accessible Rich Internet Applications}
\newacronym{ssr}{SSR}{Server-Side Rendering}
\newacronym{ssg}{SSG}{Static Site Generation}
\newacronym{ajax}{AJAX}{Asynchronous JavaScript and XML}
\newacronym{xhr}{XHR}{XML HTTP Request}
\newacronym{spa}{SPA}{Single Page Application}
\newacronym[longplural={Content Security Policies}]{csp}{CSP}{Content Security Policy}
\newacronym{xss}{XSS}{Cross-Site Scripting}
\newacronym{wcag}{WCAG}{Web Content Accessibility Guidelines}
\newacronym{cdn}{CDN}{Content Delivery Network}
\newacronym{rop}{ROP}{Return Oriented Programming}

\lstdefinestyle{html}{
  basicstyle=\footnotesize\fontfamily{pxtt}\selectfont,
  keywordstyle=\bfseries,
  showstringspaces=false,
  frame=tb,
}

\lstdefinestyle{javascript}{
  basicstyle=\footnotesize\fontfamily{pxtt}\selectfont,
  keywordstyle=\bfseries,
  showstringspaces=false,
  frame=tb,
}

\newcommand{\SI}[2]{%
  #1\,#2%
}

\newcommand{\javascript}{Java\-Script}
\newcommand{\tranco}{{\sc Tranco}}

\newcommand{\posthtml}{PostHTML}
\newcommand{\pluginname}{\jsrehab{} plugin}

\newcommand{\firefox}{Firefox}
\newcommand{\talkback}{TalkBack}
\newcommand{\android}{Android}
\newcommand{\chrome}{Chrome}
\newcommand{\curl}{cURL}
\newcommand{\publicwww}{PublicWWW}

\newcommand{\bootstrap}{Bootstrap}
\newcommand{\zurbfoundation}{Foundation}

\newcommand{\webpack}{Webpack}
\newcommand{\rollupjs}{rollup.js}
\newcommand{\expressjs}{Express}

\newcommand{\jscleaner}{JSCleaner}
\newcommand{\jsrehab}{\textsc{JSRehab}}
\newcommand{\noscript}[0]{noscript}

\newcommand{\htmltag}[1]{\texttt{<#1>}}

\newcommand{\htmlattr}[1]{\texttt{#1}}

\newcommand{\csspseudoclass}[1]{\texttt{:#1}}

\newcommand{\bootstrapshare}{\SI{20.7}{\%}}
\newcommand{\samplesize}{\SI{100}{webpages}}
\newcommand{\medianoverhead}{\SI{5}{\%}}
\newcommand{\transformationdurationmedian}{\SI{125}{ms}}

\newcommand{\crawledurlcount}{\SI{21,341}{pages}}
\newcommand{\bsanywithcomponenturlcount}{\SI{3,291}{pages}}
\newcommand{\bsfourwithcomponenturlcount}{\SI{1,372}{pages}}
\newcommand{\manuallyevaluatedcount}{\SI{100}{pages}}

\newcommand{\manualworkingpages}{\SI{79}{pages}}
\newcommand{\manualmiscerrors}{\SI{19}{pages}}
\newcommand{\manualrealissues}{\SI{2}{pages}}

\newcommand{\lowenergysavings}{\SI{5}{\%}}

\newcommand{\pageconsocount}{\SI{31}{pages}}

\newcommand{\repourl}{\url{https://gitlab.inria.fr/jsrehab}}

\copyrightyear{2022}
\acmYear{2022}
\setcopyright{none}\acmConference[WWW '22 Companion]{Companion Proceedings of the Web Conference 2022}{April 25--29, 2022}{Virtual Event, Lyon, France}
\acmBooktitle{Companion Proceedings of the Web Conference 2022 (WWW '22 Companion), April 25--29, 2022, Virtual Event, Lyon, France}
\acmDOI{}
\acmISBN{}

\title{\jsrehab{}: Weaning Common~Web~Interface~Components from~JavaScript~Addiction}
\date{}

\author{Romain Fouquet}
\orcid{0000-0003-0369-7459}
\affiliation{%
  \institution{Inria / Univ.\,Lille}
  \postcode{F-59000}
  \city{Lille}
  \country{France}}
\author{Pierre Laperdrix}
\orcid{0000-0001-6901-3596}
\affiliation{%
  \institution{CNRS / Univ.\,Lille / Inria}
  \city{Lille}
  \country{France}}
\author{Romain Rouvoy}
\orcid{0000-0003-1771-8791}
\affiliation{%
  \institution{Univ.\,Lille / Inria}
  \city{Lille}
  \country{France}}



\begin{document}
  \begin{abstract}
  Leveraging \acrfull{js} for \acrfull{ui} interactivity has been the norm on the web for many years.
  Yet, using \acrshort{js} increases bandwidth and battery consumption as scripts need to be downloaded and processed by the browser.
  Plus, client-side \acrshort{js} may expose visitors to security vulnerabilities such as \acrfull{xss}.

  This paper introduces a new server-side plugin, called \jsrehab{}, that automatically rewrites common web interface components by alternatives that do not require any \acrfull{js}.
  The main objective of \jsrehab{} is to drastically reduce---and ultimately remove---the inclusion of \acrshort{js} in a web page to improve its responsiveness and consume less resources.
  We report on our implementation of \jsrehab{} for \bootstrap{}, the most popular \acrshort{ui} framework by far, and evaluate it on a corpus of~\samplesize{}.
  We show through manual validation that it is indeed possible to lower the dependencies of pages on \acrshort{js} while keeping intact its interactivity and accessibility.
  We observe that \jsrehab{} brings energy savings of at least \lowenergysavings{} for the majority of web pages on the tested devices, while introducing a median on-the-wire overhead of only \medianoverhead{} to the \acrshort{html} payload.
\end{abstract}

  \maketitle

  \section{Introduction}
Since its introduction in 1995, \acrfull{js} has been widely adopted by websites and is now prevalent on the web.
From web pages, embedding a few event listeners, to complex \glspl{spa}, whose interface entirely rely on \acrshort{js}, almost every website uses \acrshort{js} in one way or another~\cite{w3techs-js-usage}.

Existing tools and previous works can reduce the amount of \acrshort{js} shipped to clients, \emph{e.g.}, by removing useless code.
In particular, major bundlers---such as \webpack{}~\cite{webpack-tree-shaking} and \rollupjs{}~\cite{rollupjs-tree-shaking}---implement tree-shaking techniques to remove dead \acrshort{js} code by analyzing the import graph.
Such dead code is widespread as it is common to include large third-party \acrshort{js} libraries and to only use some of the features they provide.
Chaqfeh~\emph{et~al.}\ have investigated the classification of scripts as \emph{essential} or \emph{non-essential} as part of a rewriting proxy, with the aim of removing non-essential \acrshort{js} to improve performance on clients, especially on mobile devices~\cite{DBLP:conf/www/ChaqfehZHS20}.
However, automatically replacing parts of \acrshort{js} implementations by \noscript{} alternatives remains an unexplored strategy.

In this paper, we show that most \acrfull{ui} components provided by popular frameworks, such as \bootstrap{}~\cite{bootstrap-homepage}, can be automatically replaced by \noscript{} alternatives.
In particular, we introduce an automated \acrshort{html} rewriting technique, named \jsrehab{}, to replace these \acrshort{js} components by their \noscript{} alternatives, hence reducing the dependency on client-side \acrshort{js}, even removing it in some cases.
This contribution facilitates the deployment of stricter \glspl{csp}, enabling to entirely forbid client-side scripting if the page makes no other use of \acrshort{js}, hence dramatically improving client-side security.
Reducing the amount of client-side \acrshort{js} can also bring performance improvements and energy savings, by optimizing the amount of data transferred and of scripts processed by the browser, which can be significant on low-end mobile devices~\cite{DBLP:conf/www/ChaqfehZHS20}.
The contributions covered by this paper include:
\begin{compactenum}
  \item introducing a stateful component abstraction to implement \noscript{} alternatives,
  \item reporting on the implementation of a \noscript{} alternative generator, currently targeting the most popular \acrshort{ui} framework, \bootstrap{},
  \item evaluating the payload overhead of these \noscript{} alternatives,
  \item measuring the energy savings on mobile devices, and
  \item manually validating these \noscript{} alternatives on a corpus of \samplesize{}.
\end{compactenum}

  \section{Rewriting HTML pages with Noscript Alternatives}\label{sec:noscript-fallback-method}
This section introduces the principles underlying \noscript{} alternatives and their automated generation using \jsrehab{}.

\subsection{Introducing Noscript Alternatives}\label{subsec:noscript-alternatives}
We define a \noscript{} alternative as a web structure that implements an interactive behavior equivalent to the \acrshort{js} component it replaces.
This structure may combine \acrshort{html} and \acrshort{css} constructs to implement the expected interactivity with no single line of \acrshort{js} executed on the page.
It should be noted that, despite the naming similarity, \noscript{} alternatives are not necessarily embedded as \texttt{<noscript>} tags---which are only interpreted by the browser when \acrshort{js} is disabled---but can be set up to be rendered by any user.

When rewriting a \acrshort{ui} component as a \noscript{} alternative, one needs to store and to update the component's \textit{state} by only leveraging \acrshort{html} and \acrshort{css} constructs.
For interactive components, this state can encode for example an \textsf{opened}/\textsf{closed} menu, \textsf{clicked}/\textsf{focused} button or  \textsf{checked}/\textsf{unchecked} checkbox.
Without this state, the page cannot react to user interactions, as no component will record the changes triggered by the associated events.

To deal with this challenge, we leverage the \emph{checkbox hack}~\cite{seddon-css-checkbox, css-tricks-checkbox-hack}, making it possible to record any component state by hiding a checkbox underneath.
Interestingly, checkboxes can be natively toggled as \textsf{checked} or \textsf{unchecked} and, even if they are invisible, users can indirectly update them, offering a perfect candidate for implementing our \noscript{} alternatives.
\autoref{fig:checkbox-dropdown} illustrates such an example of a checkbox hack implementing a ``dropdown button'' label that is visible to the user, while the \texttt{\#chkbox0} element is kept hidden.
The menu is not visible as its current style is set to \texttt{display:none} but, as soon as the user clicks on the label, the state of \texttt{\#chkbox0} is toggled to \texttt{checked} and the menu becomes visible with \texttt{display:block}.

\begin{lstlisting}[
  style=html,
  language=HTML,
  caption={Noscript alternative of a dropdown button},
  label={fig:checkbox-dropdown}
]
  <div class="dropdown">
      <!-- dropdown state -->
      <input id="chkbox0" type="checkbox"
        style="position:fixed;opacity:0">
      <!-- dropdown label -->
      <label for="chkbox0">Dropdown button</label>
      <!-- dropdown menu -->
      <ul class="dropdown-menu">
          <li><a href="/item0">Item #0</a></li>
          <li><a href="/item1">Item #1</a></li>
      </ul>
  </div>
  <style>
      .dropdown-menu { display: none; }
      .dropdown #chkbox0:checked ~ .dropdown-menu {
          display: block;
      }
  </style>
\end{lstlisting}

More generally, implementing \noscript{} alternatives requires to identify:
\begin{inparaenum}[(a)]
  \item \acrshort{html} elements that are stateful and that the user can interact with, and
  \item \acrshort{css} selectors that can access the state of these elements.
\end{inparaenum}
By combining both, one can implement \acrshort{ui} components without \acrshort{js}.
We studied the \emph{CSS Selectors} specification~\cite{css-selectors-v4-spec-overview} and derived the set of pseudo-classes that can be used to access the state of \acrshort{html} elements and other mechanisms without \acrshort{js}.
\autoref{tab:css-selectors-elements} reports on the mechanisms that we leveraged in \jsrehab{}:
\begin{inparaenum}[\em (1)]
   \item \emph{checkboxes} to record boolean states,
   \item \emph{radio buttons} to store mutually exclusive boolean states,
   \item \emph{target links} to help page navigation, and
   \item \emph{hover/focus} to notify a component of page-level user interactions.
\end{inparaenum}

\begin{table}
  \centering
  \Description{Six CSS selectors usable to access element state}
  \caption{\acrshort{css} selectors \& elements/mechanisms whose state can be accessed}\label{tab:css-selectors-elements}
  \begin{tabular}{l l}
    \toprule
    \acrshort{css} selector                              & \acrshort{html} element/mechanism\\
    \midrule
    \csspseudoclass{checked}                             & \texttt{<input type="checkbox">}\\
    \csspseudoclass{checked}                             & \texttt{<input type="radio">}\\
    \csspseudoclass{target}                              & Current document's \acrshort{url} fragment\\
    \csspseudoclass{focus}/\csspseudoclass{focus-within} & Document focus\\
    \csspseudoclass{hover}                               & Cursor position\\
    \bottomrule
  \end{tabular}
\end{table}

\subsection{Rewriting UI Components with \jsrehab{}}\label{subsec:replacing-components-with-noscript-alternatives}
\subsubsection{Designing \noscript{} alternatives}
The design of a \noscript{} alternative for any \acrshort{ui} component requires to analyze:
\begin{inparaenum}[\em (1)]
  \item the \acrshort{ui} framework's documentation and source code to identify the purpose and behavior of the component,
  \item the best strategy in \autoref{tab:css-selectors-elements} to store the component's state.
\end{inparaenum}
While this approach can be applied to any \acrshort{ui} framework, inferring the exact transformation cannot be automated: each framework includes specificities, which may be encoded in a very specific way with a different architecture and corner-cases.
This requires each transformation to be manually crafted in order to make sure that everything is appropriately tailored for the targeted \acrshort{ui} framework.

\subsubsection{Generating \noscript{} alternatives}
Even though the \emph{checkbox hack} has been well-known for more than a decade~\cite{seddon-css-checkbox}, it is not widely used on the web.
This can be explained by several factors.
Firstly, the implementation of \noscript{} alternatives cannot be factored out and requires to be repeated for each component instance, making it a relatively verbose and error-prone solution when hand-writing the required \acrshort{html} and \acrshort{css}.
Moreover, this first point particularly stands out when compared to interface component frameworks---such as \bootstrap{}~\cite{bootstrap-homepage} or \zurbfoundation{}~\cite{zurb-foundation-homepage}---which only require the web developer to add a few classes and attributes to the \acrshort{ui} components to enable \acrshort{js} behaviors.
Secondly, some \noscript{} alternatives may suffer from corner-cases and unexpected behaviors, making their implementation subtle, which is worsened by the first point, as their implementations cannot be factored out.

However, we believe that the above limitations can be addressed by leveraging an \acrshort{html} preprocessor, which makes it possible to factor out the \noscript{} implementations as a transform function, thus providing polished and accessible \noscript{} alternatives.
To the best of our knowledge, this work is the first to propose using \acrshort{html} rewriting rules to automatically generate \noscript{} alternatives to common web interface components.
Using the technique detailed in \autoref{subsec:replacing-components-with-noscript-alternatives}, we succeeded to implement \noscript{} alternatives for almost all \bootstrap{} components: the list of \bootstrap{} components{} and associated \noscript{} mechanisms leveraged to replace them can be found in \autoref{tab:bootstrap-components} and the \pluginname{} repository\footnote{\repourl{}} contains detailed documentation about each component.

We opted to use an \acrshort{html} preprocessor called \posthtml{}~\cite{posthtml-repo} and create our own \pluginname{} to carry out the transformation.
By using existing transformation tooling, we also benefit from its integration into the web ecosystem, including bundlers, such as \webpack{}~\cite{webpack-homepage} and \rollupjs{}~\cite{rollupjs-homepage}, and web server frameworks, such as \expressjs{}~\cite{expressjs-homepage}.
We can also generate \noscript{} alternatives in both \gls{ssg} and \gls{ssr} contexts, by injecting them only once in the former or whenever the page is rendered in the latter case.

\begin{table}[H]
  \centering
  \caption[]{Bootstrap components having a built-in \javascript{} behavior~\cite{bootstrap5-components};
  \Description{Fourteen Bootstrap components and noscript mechanisms used to replace twelve of them}
}\label{tab:bootstrap-components}
  \begin{tabular}{l l l}
    \toprule
    \makecell[l]{Bootstrap\\component\\(latest version)} & \makecell[l]{\noscript{}\\alter-\\native} & \makecell[l]{Noscript mechanism(s) used}\\
    \midrule
    Accordion (5)        & Yes & \htmltag{input type="radio"}\\
    Affix (3)            & Yes & \texttt{position: sticky}\\
    Alerts (5)           & Yes & \htmltag{input type="checkbox"}\\
    Carousel (5)        & Yes & \htmltag{input type="radio"}\\
    Collapse (5)        & Yes & \htmltag{input type="checkbox"}\\
    Dropdowns (5)       & Yes & \htmltag{input type="checkbox"}\\
    Modal (5)           & Yes & \htmltag{input type="checkbox"}\\
    Navs \& tabs (5)    & Yes & \htmltag{input type="radio"}/\csspseudoclass{target}\\
    Offcanvas (5)       & Yes & \htmltag{input type="checkbox"}\\
    Popovers (5)        & Yes & \htmltag{input type="checkbox"}\\
    Scrollspy (5)       & No & \emph{no access to viewport in \acrshort{css}}\\
    Toasts (5)          & Yes & \htmltag{input type="checkbox"}\\
    Tooltips (5)        & Yes & \csspseudoclass{hover}/\csspseudoclass{focus}\\
    Typeahead (2)       & No & \emph{cannot replicate autocompletion}\\
    \bottomrule
  \end{tabular}
\end{table}

\subsection{About Accessibility Challenges}
Web accessibility is the practice of ensuring that there are no direct barriers to interact with a website for people with specific disabilities.
In the case of the \pluginname{}, we have to ensure that our \noscript{} alternatives are not making the web harder to browse, by providing at least as good accessibility than the replaced frameworks, and to comply with legal requirements.
Most countries having laws mandating accessibility for certain websites rely on the \acrshort{wcag}~\cite{w3c-a11y-policies} which do not specify implementation details, only high-level requirements, such as the Success Criterion~2.1.1 Keyboard~\cite{wcag21-211}, indicating only that the page must be operable with a keyboard with no time-sensitive input.

As browsers already implement accessibility for standard \acrshort{html} elements---\emph{e.g.}, spacebar toggles checkboxes, and their change of state is properly announced by screen readers---\noscript{} alternatives are accessible by default, making redundant \acrshort{waiaria} state attributes~\cite{wai-aria-stateattr}---such as \htmlattr{aria-checked}---which could not be toggled without \acrshort{js}.

  \section{Validation Methodology}
This section covers the methodology we applied to validate the \noscript{} alternatives generated by \jsrehab{} for \bootstrap{} components.

\subsection{Validation Corpus Selection}
As we focused on \bootstrap{}, we built a corpus of web pages that use this \acrshort{ui} framework, so that we can
\begin{inparaenum}
  \item obtain detailed statistics about its usage, and
  \item test and validate \jsrehab{} on them.
\end{inparaenum}
We crawled the Top~\SI{10}{k} domains from the \tranco{} list~\cite{tranco-list-homepage} with the following strategy: from the landing page, up to three \acrshortpl{url} were extracted as a random sample of \htmltag{a} tags \htmlattr{href} \acrshortpl{url} sharing the same origin as the document's \acrshort{url}, but with a different path.
Thus, our measurements were collected from up to four pages per domain: the landing page and up to three internal pages.
For each web page, we detect the use of \bootstrap{} through a combination of
\begin{inparaenum}
  \item detecting the version number exposed in the global object,
  \item parsing the source code to get the version number from a banner comment, and
  \item using custom heuristics when other methods cannot work.
\end{inparaenum}
We observed that \bootstrap{}'s \acrshort{js} is used on \bootstrapshare{} of crawled pages and that its adoption is uniformly distributed across websites ranking, except for the very best ranked websites.

We also measured the popularity of \bootstrap{}'s components by parsing the page's \acrshort{html} and we conclude that the \textsf{collapse}, \textsf{dropdown}, and \textsf{modal} components were the most common, by far, being found on \SI{53}{\%} of pages using \bootstrap{}, \SI{40}{\%}, and \SI{31}{\%}, respectively, while all other components are found on less than \SI{10}{\%} of pages using \bootstrap{}.

Among the \crawledurlcount{} we crawled, \bsanywithcomponenturlcount{} were using \bootstrap{}'s \acrshort{js} and included at least one component in the crawled page.
Among these, \bsfourwithcomponenturlcount{} were using \bootstrap{}~4 or~5.
Validation is split into two parts: measuring rewriting statistics over the whole sample of \bsfourwithcomponenturlcount{}, and empirically evaluating the interactivity and accessibility of \noscript{} alternatives on desktop and mobile devices on a sample of \manuallyevaluatedcount{}.

\vspace{-0.4ex}
\subsection{Validation Setup}
\begin{figure*}
  \centering
  \includegraphics[width=.75\linewidth]{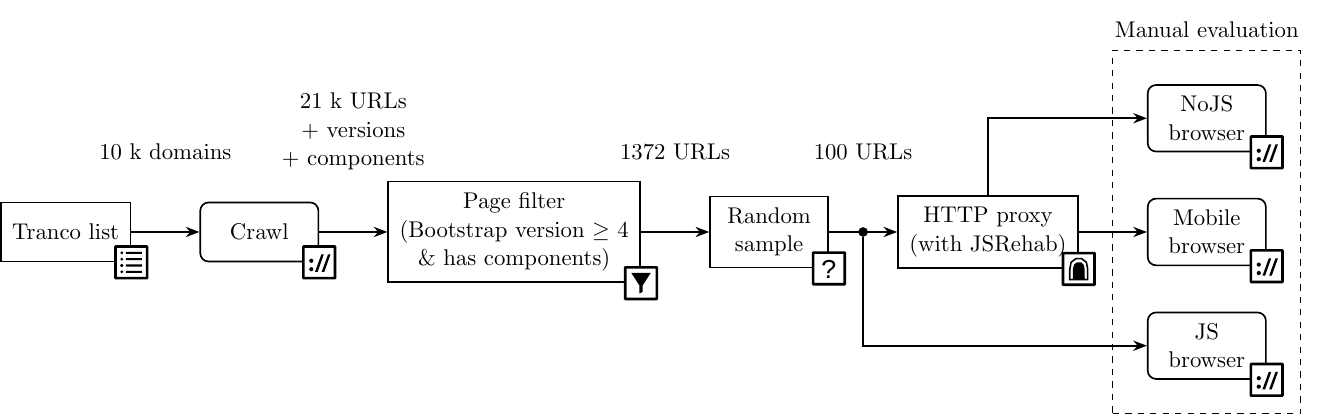}
  \Description{Dataflow diagram of the validation setup, from the Tranco list to manual evaluation in browsers}
  \vspace{-3ex}
  \caption{Dataflow diagram of our empirical evaluation setup}\label{fig:evaluation-data-flow}
\end{figure*}

Since it would not have been possible to deploy our solution on production web servers or as a static site generator for testing, we rather chose to implement an \acrshort{http} rewriting proxy, which applies the \acrshort{html} transformation on the fly, whenever a page is requested, see \autoref{fig:evaluation-data-flow}.

\subsubsection{Rewriting Statistics}
To collect statistics about all web pages from the validation set, the list of \bsfourwithcomponenturlcount{}  is passed to \curl{} configured to use the \acrshort{http} proxy and the \firefox{} user-agent header, as some websites reject requests with no user-agent.

The \acrshort{http} proxy saves the transformation duration and the original and transformed sizes of the compressed \acrshort{http} response body, using the same compression method as used by the website (gzip or brotli).

\subsubsection{Empirical Validation}
The empirical validation is achieved by visiting the same \acrshort{url} three times: once in a control browser on desktop and with \acrshort{js} enabled, a second time in a browser with \acrshort{js} disabled by default and connected to the \acrshort{http} proxy, and a third time with a mobile browser, as shown in \autoref{tab:testing-browser-configurations}.
Two viewport widths are tested as webpages often include a hamburger menu that only appears on mobile, leveraging \acrshort{css} media queries for responsive design.

We chose to disable \acrshort{js} by default in the second browser, so that the original \bootstrap{}'s \acrshort{js} and other custom \acrshort{js} added on the page for component interactivity do not interfere with the \noscript{} alternatives; we only temporarily enable \acrshort{js} to be able to access some hidden components on the page.
This, however, makes it impossible to compare the page load time or the time-to-interactive between the pages with and without \noscript{} alternatives, as it could not be isolated from the mere \acrshort{js} blocking.

\begin{table*}
  \centering
  \caption{Browser configurations used for our empirical evaluation}\label{tab:testing-browser-configurations}
  \Description{The three browser configurations used for the evaluation}
  \begin{tabular}{l l l l l c}
    \toprule
    Browser & Device & \makecell[l]{Window width (px)} & Input device & \makecell[l]{Screenreader} & \acrshort{js}\\
    \midrule
    \firefox{}~93.0 & Desktop (Debian) & 1280 & Keyboard & \multirow{1}{*}{No} & Yes / No\\
    \firefox{}~93.0 & Desktop (Debian) & 720 (responsive mode) & Keyboard & \multirow{1}{*}{No} & Yes / No\\
    \firefox{}~92.0.1 & Mobile (\android{}) & 1440 & Touchscreen & Yes (\talkback{} 2021-04) & No\\
    \bottomrule
  \end{tabular}
\end{table*}

\begin{figure}
  \centering
  \includegraphics[width=.9\linewidth]{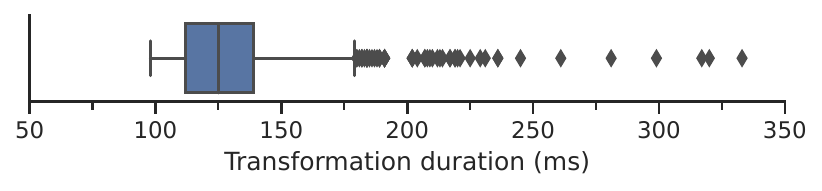}
  \Description[Boxplot of the transformation durations]{Boxplot of the transformation durations, the median is 125 ms}
  \caption{Distribution of transformation durations on the corpus of tested pages}\label{fig:transformation-duration-dist}
\end{figure}

\subsection{Compatibility Validation}
After loading the web page in the browser, we manually validated the interactivity and accessibility of the \noscript{} alternatives by checking the following features:
\begin{compactitem}
  \item on desktop devices:
    \begin{compactitem}
      \item \noscript{} alternatives can be activated with a pointing device,
      \item tab-controlled focus behave properly, and
      \item \noscript{} alternatives can be activated with spacebar/arrow keys.
    \end{compactitem}
  \item on mobile devices:
    \begin{compactitem}
      \item \noscript{} alternatives can be activated with a touch device,
      \item \noscript{} alternatives can be focused using screenreader navigation,
      \item \noscript{} alternatives can be activated with screenreader navigation, and
      \item screenreader speech announces \noscript{} alternatives appropriately (providing understandable navigation).
    \end{compactitem}
\end{compactitem}

For each web page, our testing protocol is as follows:
\begin{compactenum}
  \item we verify that the original components on the page are working with \acrshort{js}, but are unresponsive without it,
  \item we validate the aforementioned criteria on a modified page on both desktop and mobile devices with all the \noscript{} alternatives included.
\end{compactenum}
To ease the validation, the \acrshort{http} proxy highlights the \noscript{} alternatives so that they are easier to locate and test.
For components hidden by default, especially modals, which could not be shown as \acrshort{js} is disabled, some additional effort is made on desktop to make them appear, so that the \noscript{} alternatives can be assessed.

Finally, only \bootstrap{} components---originally using \bootstrap{}'s \acrshort{js}---are validated, while other components of the page are left untested.

  \section{Empirical Results}
This section reports on the results we obtained by evaluating \pluginname{} on the validation corpus.

\subsection{Rewriting Statistics}
\begin{figure}
  \centering
  \includegraphics[width=.9\linewidth]{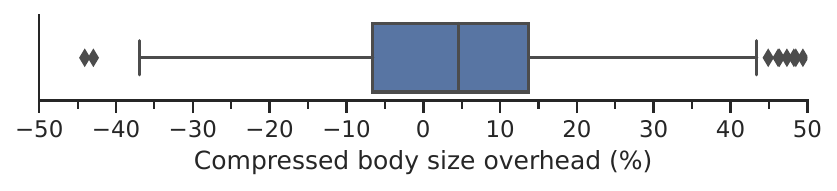}
  \Description[Boxplot of the compressed body overhead]{Boxplot of the compressed body overhead, the median is 4.6 \%}
  \caption{Distribution of the compressed body size overhead on the set of tested pages (some outliers are omitted to improve readability)}\label{fig:compressed-body-overhead-dist}
\end{figure}

As \noscript{} alternatives are injected in the \acrshort{html} document, they increase its size, the difference depending on the type and the number of components included in the page.
However, as depicted in \autoref{fig:compressed-body-overhead-dist}, the resulting overhead on the \emph{compressed} body of the \acrshort{http} response containing the \acrshort{html} document is extremely low, with a median overhead of \medianoverhead{}; the overhead is lower than \SI{15}{\%} for more than \SI{75}{\%} of tested pages.

The distribution of the rewriting delay, measured by the \acrshort{http} proxy running on a high-end laptop for testing, can be found in \autoref{fig:transformation-duration-dist}.
The median rewriting delay is lower than \transformationdurationmedian{} on the sample of tested pages.
The rewriting delay mostly depends on the number of \acrshort{html} nodes in the page, as generating the \noscript{} alternatives requires multiple tree traversals.

\subsection{Manual Validation}
Among the \manuallyevaluatedcount{} manually analyzed, \manualworkingpages{} were testable, and all \noscript{} alternatives were working in compliance with the criteria defined in the testing methodology, \manualmiscerrors{} had various issues preventing complete testing, and issues effectively due to \noscript{} alternatives were only found on \manualrealissues{}, as detailed in \autoref{tab:evaluation-results}.

Pages not being fully testable include error pages, which were not the intended pages, \glspl{spa} leaving a blank page when \acrshort{js} is disabled, pages with components dynamically added, which thus cannot be detected when processing the \acrshort{html} document, and pages that were updated between the initial crawl and the validation.

The two pages presenting issues included unconventionally used \bootstrap{} components.
One of them used collapse components with \htmlattr{data-parent} attributes on several different buttons of the web interface.
This attribute is intended to be used to build accordions~\cite{bootstrap4-accordion}, which are supported by the \pluginname{}; however, this page uses them to make the page menus mutually exclusive so that at most one is open at any time.
The other page only partially implements the markup to make footer section headers collapse buttons on mobile, the \pluginname{} produces collapse buttons for these headers that are also enabled on desktop.

For all other tested pages, the \pluginname{} produced effective \noscript{} alternatives to original components.

\begin{table*}
  \centering
  \Description{Results of manual in-browser evaluation}
  \caption{Summary of our empirical evaluation observations on a sample of \manuallyevaluatedcount{}}\label{tab:evaluation-results}
  \begin{tabular}{l r}
    \toprule
    Observed behavior & Count\\
    \midrule
    Web page is fully interactive and all \noscript{} alternatives are behaving correctly & \manualworkingpages{}\\
    No component found in the page with \acrshort{js}: the page likely changed between the initial crawl and the validation & 6~pages\\
    Error pages on web pages with \acrshort{js} enabled (different from the initial crawl) & 4~pages\\
    Buggy pages due to inappropriate usage of original \bootstrap{} & 4~pages\\
    \glspl{spa} or components dynamically added by \acrshort{js} & 3~pages\\
    Custom component styling that cannot be triggered by \noscript{} alternatives and cannot be manually bypassed for testing & 2~pages\\
    Web page is interactive, but some \noscript{} alternatives are misbehaving & \manualrealissues{}\\
    \bottomrule
  \end{tabular}
\end{table*}

\subsection{Preliminary Measurements of Consumption}
To complement the rewriting statistics, we measured the energy consumption on mobile devices of a sample of web pages that use \bootstrap{}'s \acrshort{js}.
Using \android{}'s built-in utility \texttt{dumpsys batterystats}, which can query the device consumption on a per-app basis, we compared the consumption of \chrome{} loading the unmodified page with a modified version where \bootstrap{}'s \acrshort{js} is blocked and \jsrehab{} is used to preserve page functionality.
We focused on \chrome{} as it is by far the most popular browser on \android{}~\cite{statcounter-mobile-browser-market-share}.

We use a rewriting proxy similar to the one presented above, serving for each requested page:
\begin{inparaenum}[\em (1)]
  \item the original version with \acrshort{js} and
  \item the version rewritten with \jsrehab{} and blocking \bootstrap{}'s \acrshort{js}, while still allowing other \acrshort{js} to load and execute.
\end{inparaenum}
As isolating and blocking \bootstrap{}'s \acrshort{js} is not possible on every page, we focused on the top \pageconsocount{} that include a file \texttt{bootstrap.min.js}, according to \publicwww{}~\cite{publicwww-bootstrapjs}.
The proxy is configured to prevent all \acrshort{http} caching from the browser and injects a \texttt{Refresh=5} \acrshort{http} header, which forces the page to reload every \SI{5}{s}.
The energy consumption for each version of the requested web page is measured for \SI{180}{s}, effectively averaging the measurements over \SI{36}{page~loads}, while we made sure that each page was able to load within the \SI{5}{s} timeframe.
The reported measurements have been performed on two low-to-medium-end phones, the Archos~50 Platinum~4G and the Moto~Z, respectively running \chrome{}~50 and \chrome{}~96.

\begin{figure}
  \centering
  \includegraphics[width=.9\linewidth]{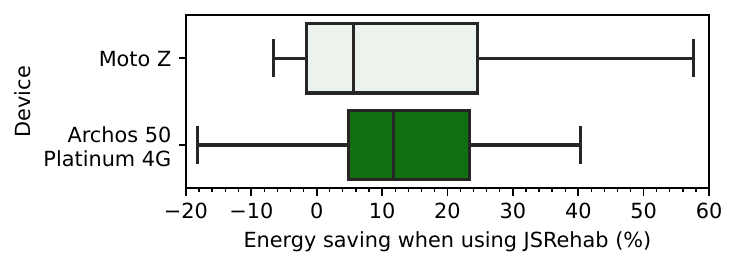}
  \Description[Boxplot of the energy savings on two mobile phones]{Boxplot of the energy savings on two mobile phones, the medians are 4.8 \% and 11\%}
  \caption{Energy savings of mobile devices when loading web pages without \bootstrap{}'s \acrshort{js} and rewritten with \jsrehab{}; outliers are excluded for readability}\label{fig:consumtion-measurements}
\end{figure}

As depicted in \autoref{fig:consumtion-measurements}, one can observe that replacing \bootstrap{}'s \acrshort{js} with \jsrehab{} enabled significant energy savings on many web pages on the tested devices; websites operators should evaluate on a case-by-case basis the exact benefit on their own website.

  \section{Discussion}
\subsection{Expected Benefits}
\subsubsection{Improving security and privacy}
Deploying the \pluginname{} makes it possible to remove the \bootstrap{} dependency, which can then be removed from allowed \acrshort{js} sources in the \gls{csp}, hence contributing to reducing the attack surface.
If the web page makes no other use of \acrshort{js}, the execution of \acrshort{js} can even be forbidden in the page by adopting a strict \gls{csp} directive, further mitigating the risk of \gls{xss}.
Website owners may thus be incentivized to further reduce the amount of \acrshort{js} included in their web pages, as one of the key usage of \acrshort{js} directly benefiting the user---component interactivity---has been substituted.

Another strong incentive to use \jsrehab{} is that it protects a website from yet-to-be-discovered vulnerabilities.
At the time of development, a developer can integrate the latest available version of a \acrshort{ui} library that may be assumed as safe.
Then, weeks later, a vulnerability can be discovered, hence requiring the dependency to be updated.
With \jsrehab{}, a website is protected as the \acrshort{js} code is simply not there.
This problem is widespread as a lot of the crawled websites were using outdated versions of \bootstrap{}.
This lack of security fixes can open users to undesired security problems.

\subsubsection{Improving performance}
Replacing \acrshort{js} components with their \noscript{} alternatives can also bring performance improvements.
Indeed, \noscript{} alternatives adding only a median \medianoverhead{} overhead, and since \bootstrap{}'s \acrshort{js} is not needed if the page only includes the supported components, the amount of data transferred on the wire is reduced, leading to faster page loads.
This point is even more significant as major browsers have implemented \acrshort{http} cache partitioning~\cite{chrome-cache-partitioning, firefox-state-partitioning}, preventing \bootstrap{}'s \acrshort{js} to be reused between websites when the same version was loaded from a \acrshort{cdn}.

Moreover, the processing burden on the client is reduced as the browser does not need to parse and execute the additional \acrshort{js} (\bootstrap{}~5's \acrshort{js} weights \SI{60}{kB} minified but uncompressed), which can be significant, especially on mobile devices~\cite{DBLP:conf/www/ChaqfehZHS20} and can extend device lifespan.
The time-to-interactive of web pages can also be reduced, thus improving page responsiveness.
Typically, such performance improvements are not achieved by \jscleaner{}~\cite{DBLP:conf/www/ChaqfehZHS20}, as component interactivity scripts would be considered as \emph{essential}.

\subsection{Ease of Adoption}
When using \bootstrap{} components as intended, the \pluginname{} produces effective \noscript{} alternatives with almost no configuration.
It only needs to be provided with the stylesheets used by the page, so that it can generate matching styling.
As \posthtml{} is already integrated into various bundlers and web server frameworks, such as \webpack{}, \rollupjs{} or \expressjs{}, it is straightforward to adopt \jsrehab{} in an existing project, with no change in tooling; the \jsrehab{} repository contains an example of configuration file for \webpack{}.

Furthermore, as \noscript{} alternatives are generated for each page separately, it is possible to progressively transition to using the \pluginname{} and enforcing a stricter \acrshort{csp}.

\glsreset{ssg}
\glsreset{ssr}
With a median transformation delay of \transformationdurationmedian{}, \jsrehab{} performance is compatible with a \gls{ssg} setup, where pages are rendered once, then served as part of a static site.
Depending on the website expectations, it may currently be too slow for a \gls{ssr} context, where pages are rendered on the web server upon requests.
The \pluginname{} would need to be further optimized for this context or rewritten in a programming language more suited to string processing.

\subsection{Beyond \bootstrap{}}
We specifically demonstrated the generation of \noscript{} alternatives for replacing components from the \bootstrap{} framework, but this technique can be leveraged for other component frameworks as well, be them public or in-house, to factor out \noscript{} alternative implementation and make them easier to use.
As many \acrshort{ui} frameworks share a large set of components~\cite{zurb-foundation-homepage, semanticui-homepage}, porting the \pluginname{} would mostly require updating the class names used as component identifiers, which are specific to each framework.

Other types of components, not implemented by \bootstrap{}, could also be automatically replaced with \noscript{} alternatives, including sortable tables, image lightboxes, and data plots.
Other components could be implemented without \acrshort{js} if browsers were to implement \acrshort{css} pseudo-classes such as \csspseudoclass{has()}~\cite{csswg-has-pseudoclass} or \csspseudoclass{in-viewport}~\cite{webwewant-in-viewport}.

  \section{Conclusion}
In this paper, we introduced a server-side technique to automatically replace common web interface components implemented by \acrshort{ui} frameworks with \noscript{} alternatives.
We implemented this technique as a set of \acrshort{html} rewriting rules that generate \noscript{} alternatives for \bootstrap{} and we discussed the key benefits and current limitations of our contribution.
We also validated these \noscript{} alternatives on a corpus of~\samplesize{}, and we observed that they deliver convincing alternatives by assessing their interactivity and accessibility from both desktop and mobile devices, while introducing only minimal overhead on the compressed \acrshort{html} document and that they enable energy savings.

  \bibliographystyle{ACM-Reference-Format}
  \bibliography{paper_references}


\begin{thebibliography}{26}

%%% the Web address.

\ifx \showCODEN    \undefined \def \showCODEN     #1{\unskip}     \fi
\ifx \showDOI      \undefined \def \showDOI       #1{#1}\fi
\ifx \showISBNx    \undefined \def \showISBNx     #1{\unskip}     \fi
\ifx \showISBNxiii \undefined \def \showISBNxiii  #1{\unskip}     \fi
\ifx \showISSN     \undefined \def \showISSN      #1{\unskip}     \fi
\ifx \showLCCN     \undefined \def \showLCCN      #1{\unskip}     \fi
\ifx \shownote     \undefined \def \shownote      #1{#1}          \fi
\ifx \showarticletitle \undefined \def \showarticletitle #1{#1}   \fi
\ifx \showURL      \undefined \def \showURL       {\relax}        \fi
% invisible to TeX
\providecommand\bibfield[2]{#2}
\providecommand\bibinfo[2]{#2}
\providecommand\natexlab[1]{#1}
\providecommand\showeprint[2][]{arXiv:#2}

\bibitem[\protect\citeauthoryear{??}{pos}{2021}]%
        {posthtml-repo}
 \bibinfo{year}{2021}\natexlab{}.
\newblock \bibinfo{booktitle}{\emph{PostHTML}}.
\newblock
\urldef\tempurl%
\url{https://github.com/posthtml/posthtml}
\showURL{%
\tempurl}


\bibitem[\protect\citeauthoryear{??}{rol}{2021}]%
        {rollupjs-homepage}
 \bibinfo{year}{2021}\natexlab{}.
\newblock \bibinfo{booktitle}{\emph{rollup.js}}.
\newblock
\urldef\tempurl%
\url{https://rollupjs.org/guide/}
\showURL{%
\tempurl}


\bibitem[\protect\citeauthoryear{??}{sem}{2021}]%
        {semanticui-homepage}
 \bibinfo{year}{2021}\natexlab{}.
\newblock \bibinfo{booktitle}{\emph{Semantic UI}}.
\newblock
\urldef\tempurl%
\url{https://semantic-ui.com/}
\showURL{%
\tempurl}


\bibitem[\protect\citeauthoryear{??}{web}{2021a}]%
        {webwewant-in-viewport}
 \bibinfo{year}{2021}\natexlab{a}.
\newblock \bibinfo{booktitle}{\emph{The Web We Want --- I want a CSS
  pseudo-selector for elements that are in the viewport}}.
\newblock
\urldef\tempurl%
\url{https://webwewant.fyi/wants/63/}
\showURL{%
\tempurl}


\bibitem[\protect\citeauthoryear{??}{web}{2021b}]%
        {webpack-homepage}
 \bibinfo{year}{2021}\natexlab{b}.
\newblock \bibinfo{booktitle}{\emph{webpack}}.
\newblock
\urldef\tempurl%
\url{https://webpack.js.org/}
\showURL{%
\tempurl}


\bibitem[\protect\citeauthoryear{??}{pub}{2022}]%
        {publicwww-bootstrapjs}
 \bibinfo{year}{2022}\natexlab{}.
\newblock \bibinfo{booktitle}{\emph{PublicWWW --- "bootstrap.min.js"}}.
\newblock
\urldef\tempurl%
\url{https://publicwww.com/websites/%22bootstrap.min.js%22/}
\showURL{%
Retrieved 2022-01-24 from \tempurl}


\bibitem[\protect\citeauthoryear{??}{rol}{2022}]%
        {rollupjs-tree-shaking}
 \bibinfo{year}{2022}\natexlab{}.
\newblock \bibinfo{booktitle}{\emph{rollup.js --- Tree-Shaking}}.
\newblock
\urldef\tempurl%
\url{https://rollupjs.org/guide/en/#tree-shaking}
\showURL{%
Retrieved 2022-03-04 from \tempurl}


\bibitem[\protect\citeauthoryear{??}{web}{2022}]%
        {webpack-tree-shaking}
 \bibinfo{year}{2022}\natexlab{}.
\newblock \bibinfo{booktitle}{\emph{Webpack --- Tree Shaking}}.
\newblock
\urldef\tempurl%
\url{https://webpack.js.org/guides/tree-shaking/}
\showURL{%
Retrieved 2022-03-04 from \tempurl}


\bibitem[\protect\citeauthoryear{Chaqfeh, Zaki, Hu, and Subramanian}{Chaqfeh
  et~al\mbox{.}}{2020}]%
        {DBLP:conf/www/ChaqfehZHS20}
\bibfield{author}{\bibinfo{person}{Moumena Chaqfeh}, \bibinfo{person}{Yasir
  Zaki}, \bibinfo{person}{Jacinta Hu}, {and} \bibinfo{person}{Lakshmi
  Subramanian}.} \bibinfo{year}{2020}\natexlab{}.
\newblock \showarticletitle{JSCleaner: De-Cluttering Mobile Webpages Through
  JavaScript Cleanup}. In \bibinfo{booktitle}{\emph{{WWW} '20: The Web
  Conference 2020, Taipei, Taiwan, April 20-24, 2020}},
  \bibfield{editor}{\bibinfo{person}{Yennun Huang}, \bibinfo{person}{Irwin
  King}, \bibinfo{person}{Tie{-}Yan Liu}, {and} \bibinfo{person}{Maarten van
  Steen}} (Eds.). \bibinfo{publisher}{{ACM} / {IW3C2}},
  \bibinfo{pages}{763--773}.
\newblock
\urldef\tempurl%
\url{https://doi.org/10.1145/3366423.3380157}
\showDOI{\tempurl}


\bibitem[\protect\citeauthoryear{Coyier}{Coyier}{2020}]%
        {css-tricks-checkbox-hack}
\bibfield{author}{\bibinfo{person}{Chris Coyier}.}
  \bibinfo{year}{2020}\natexlab{}.
\newblock \bibinfo{booktitle}{\emph{The “Checkbox Hack” (and things you can
  do with it)}}.
\newblock
\urldef\tempurl%
\url{https://css-tricks.com/the-checkbox-hack/}
\showURL{%
Retrieved 2022-03-04 from \tempurl}


\bibitem[\protect\citeauthoryear{Johann~Hofmann}{Johann~Hofmann}{2021}]%
        {firefox-state-partitioning}
\bibfield{author}{\bibinfo{person}{Tim~Huang Johann~Hofmann}.}
  \bibinfo{year}{2021}\natexlab{}.
\newblock \bibinfo{booktitle}{\emph{{M}ozilla Hacks --- Introducing State
  Partitioning}}.
\newblock
\urldef\tempurl%
\url{https://hacks.mozilla.org/2021/02/introducing-state-partitioning/}
\showURL{%
Retrieved 2022-03-04 from \tempurl}


\bibitem[\protect\citeauthoryear{Kitamura}{Kitamura}{2020}]%
        {chrome-cache-partitioning}
\bibfield{author}{\bibinfo{person}{Eiji Kitamura}.}
  \bibinfo{year}{2020}\natexlab{}.
\newblock \bibinfo{booktitle}{\emph{{G}oogle Developers --- Gaining security
  and privacy by partitioning the cache}}.
\newblock
\urldef\tempurl%
\url{https://developers.google.com/web/updates/2020/10/http-cache-partitioning}
\showURL{%
Retrieved 2022-03-04 from \tempurl}


\bibitem[\protect\citeauthoryear{Pochat, Goethem, Tajalizadehkhoob,
  Korczyński, and Joosen}{Pochat et~al\mbox{.}}{2021}]%
        {tranco-list-homepage}
\bibfield{author}{\bibinfo{person}{Victor~Le Pochat}, \bibinfo{person}{Tom~Van
  Goethem}, \bibinfo{person}{Samaneh Tajalizadehkhoob}, \bibinfo{person}{Maciej
  Korczyński}, {and} \bibinfo{person}{Wouter Joosen}.}
  \bibinfo{year}{2021}\natexlab{}.
\newblock \bibinfo{booktitle}{\emph{Tranco --- A Research-Oriented Top Sites
  Ranking Hardened Against Manipulation}}.
\newblock
\urldef\tempurl%
\url{https://tranco-list.eu/}
\showURL{%
\tempurl}


\bibitem[\protect\citeauthoryear{Q-Success}{Q-Success}{2021}]%
        {w3techs-js-usage}
\bibfield{author}{\bibinfo{person}{Q-Success}.}
  \bibinfo{year}{2021}\natexlab{}.
\newblock \bibinfo{booktitle}{\emph{W3Techs --- Usage statistics of client-side
  programming languages for websites}}.
\newblock
\urldef\tempurl%
\url{https://w3techs.com/technologies/overview/client_side_language}
\showURL{%
Retrieved 2021-10-19 from \tempurl}


\bibitem[\protect\citeauthoryear{Seddon}{Seddon}{2010}]%
        {seddon-css-checkbox}
\bibfield{author}{\bibinfo{person}{Ryan Seddon}.}
  \bibinfo{year}{2010}\natexlab{}.
\newblock \bibinfo{booktitle}{\emph{Custom radio and checkbox inputs using
  CSS}}.
\newblock
\urldef\tempurl%
\url{https://ryanseddon.com/css/custom-inputs-using-css/}
\showURL{%
Retrieved 2022-03-04 from \tempurl}


\bibitem[\protect\citeauthoryear{StatCounter}{StatCounter}{2022}]%
        {statcounter-mobile-browser-market-share}
\bibfield{author}{\bibinfo{person}{StatCounter}.}
  \bibinfo{year}{2022}\natexlab{}.
\newblock \bibinfo{booktitle}{\emph{statcounter --- Mobile Browser Market Share
  Worldwide}}.
\newblock
\urldef\tempurl%
\url{https://gs.statcounter.com/browser-market-share/mobile/worldwide}
\showURL{%
Retrieved 2022-01-24 from \tempurl}


\bibitem[\protect\citeauthoryear{StrongLoop, IBM, and other~expressjs.com
  contributors}{StrongLoop et~al\mbox{.}}{2021}]%
        {expressjs-homepage}
\bibfield{author}{\bibinfo{person}{StrongLoop}, \bibinfo{person}{IBM}, {and}
  \bibinfo{person}{other~expressjs.com contributors}.}
  \bibinfo{year}{2021}\natexlab{}.
\newblock \bibinfo{booktitle}{\emph{Express - Node.js web application
  framework}}.
\newblock
\urldef\tempurl%
\url{https://expressjs.com/}
\showURL{%
\tempurl}


\bibitem[\protect\citeauthoryear{team}{team}{2021a}]%
        {bootstrap4-accordion}
\bibfield{author}{\bibinfo{person}{Bootstrap team}.}
  \bibinfo{year}{2021}\natexlab{a}.
\newblock \bibinfo{booktitle}{\emph{Bootstrap 4 --- Collapse, Accordion
  example}}.
\newblock
\urldef\tempurl%
\url{https://getbootstrap.com/docs/4.6/components/collapse/#accordion-example}
\showURL{%
Retrieved 2021-10-20 from \tempurl}


\bibitem[\protect\citeauthoryear{team}{team}{2021b}]%
        {bootstrap5-components}
\bibfield{author}{\bibinfo{person}{Bootstrap team}.}
  \bibinfo{year}{2021}\natexlab{b}.
\newblock \bibinfo{booktitle}{\emph{Bootstrap 5 --- Components}}.
\newblock
\urldef\tempurl%
\url{https://getbootstrap.com/docs/5.1/components/}
\showURL{%
Retrieved 2021-10-19 from \tempurl}


\bibitem[\protect\citeauthoryear{team}{team}{2021c}]%
        {bootstrap-homepage}
\bibfield{author}{\bibinfo{person}{Bootstrap team}.}
  \bibinfo{year}{2021}\natexlab{c}.
\newblock \bibinfo{booktitle}{\emph{Bootstrap · The most popular HTML, CSS,
  and JS library in the world.}}
\newblock
\urldef\tempurl%
\url{https://getbootstrap.com/}
\showURL{%
\tempurl}


\bibitem[\protect\citeauthoryear{W3C}{W3C}{2021a}]%
        {wai-aria-stateattr}
\bibfield{author}{\bibinfo{person}{W3C}.} \bibinfo{year}{2021}\natexlab{a}.
\newblock \bibinfo{booktitle}{\emph{Accessible Rich Internet Applications
  (WAI-ARIA) 1.3 --- WAI-ARIA States and Properties}}.
\newblock
\urldef\tempurl%
\url{https://w3c.github.io/aria/#introstates}
\showURL{%
\tempurl}


\bibitem[\protect\citeauthoryear{W3C}{W3C}{2021b}]%
        {css-selectors-v4-spec-overview}
\bibfield{author}{\bibinfo{person}{W3C}.} \bibinfo{year}{2021}\natexlab{b}.
\newblock \bibinfo{booktitle}{\emph{Selectors Level 4}}.
\newblock
\urldef\tempurl%
\url{https://drafts.csswg.org/selectors-4/#overview}
\showURL{%
\tempurl}


\bibitem[\protect\citeauthoryear{W3C}{W3C}{2021c}]%
        {csswg-has-pseudoclass}
\bibfield{author}{\bibinfo{person}{W3C}.} \bibinfo{year}{2021}\natexlab{c}.
\newblock \bibinfo{booktitle}{\emph{Selectors Level 4}}.
\newblock
\urldef\tempurl%
\url{https://drafts.csswg.org/selectors/#relational}
\showURL{%
\tempurl}


\bibitem[\protect\citeauthoryear{W3C}{W3C}{2021d}]%
        {w3c-a11y-policies}
\bibfield{author}{\bibinfo{person}{W3C}.} \bibinfo{year}{2021}\natexlab{d}.
\newblock \bibinfo{booktitle}{\emph{Web Accessibility Laws \& Policies}}.
\newblock
\urldef\tempurl%
\url{https://www.w3.org/WAI/policies/}
\showURL{%
\tempurl}


\bibitem[\protect\citeauthoryear{W3C}{W3C}{2021e}]%
        {wcag21-211}
\bibfield{author}{\bibinfo{person}{W3C}.} \bibinfo{year}{2021}\natexlab{e}.
\newblock \bibinfo{booktitle}{\emph{Web Content Accessibility Guidelines (WCAG)
  2.1 --- Success Criterion 2.1.1 Keyboard}}.
\newblock
\urldef\tempurl%
\url{https://w3c.github.io/wcag21/guidelines/#keyboard}
\showURL{%
\tempurl}


\bibitem[\protect\citeauthoryear{Yetinauts}{Yetinauts}{2021}]%
        {zurb-foundation-homepage}
\bibfield{author}{\bibinfo{person}{Foundation Yetinauts}.}
  \bibinfo{year}{2021}\natexlab{}.
\newblock \bibinfo{booktitle}{\emph{Foundation --- The most advanced responsive
  front-end framework in the world.}}
\newblock
\urldef\tempurl%
\url{https://foundation.zurb.com/}
\showURL{%
\tempurl}


\end{thebibliography}
\end{document}